\documentclass[twocolumn,aps,pra]{revtex4}
\usepackage{epsfig}
\usepackage[english]{babel}
\usepackage{latexsym}
\usepackage{graphics}
\usepackage{subfigure}
\usepackage{graphicx}
\usepackage{dcolumn}
\usepackage{amsmath}
\usepackage{hyperref}
\usepackage{amssymb}
\usepackage{color}

\begin{document}
\title{Two-Color Attosecond Chronoscope}

\author{J. N. Wu$^{1,\dag}$, J. Y. Che$^{1,\dag}$, F. B. Zhang$^{1}$, C. Chen$^{1}$, W. Y. Li$^{2,*}$, G. G. Xin$^{3}$, and Y. J. Chen$^{1,\ddag}$}

\date{\today}

\begin{abstract}

We study ionization of atoms in strong orthogonal two-color ($\omega,2\omega$) (OTC) laser fields numerically and analytically.
The calculated photoelectron momentum distribution shows two typical structures:
a rectangular-like structure  and a shoulder-like structure,
the positions of which depend on the laser parameters.
Using a strong-field model which allows us to
quantitatively evaluate the Coulomb effect, we show that these two structures
arise from attosecond response of electron inside an atom to light
 in OTC-induced photoemission.
Some simple mappings between the locations of these structures and response time are derived,
with which we are able to establish two-color attosecond chronoscope with high resolution
for timing electron emission  in OTC-based precise manipulation.

\end{abstract}

\affiliation{1.College of Physics and Information Technology, Shaan'xi Normal University, Xi'an, China\\
2.School of Mathematics and Science and Hebei Key Laboratory of Photoelectronic Information and Goe-detection Technology, Hebei GEO University, Shijiazhuang, China\\
3.School of Physics, Northwest University, Xi'an, China}
\maketitle


\emph{Introduction}.-Measurement and control of motion of electrons inside atoms and molecules
at the electronic natural time scale have attracted wide interest in recent years \cite{Krausz2009,Krausz,Maquet,Vrakking,Pazourek,Vos}.
Relevant ultrafast-probing experiments are based on strong laser-matter interaction and generally
use the simple physical picture given by the so-called simple-man (SM) model \cite{Corkum}.
When atoms and molecules are exposed to strong laser fields, the valence electron of the target can escape
from the laser-Coulomb-formed barrier
 through tunneling \cite{Keldysh}.
The tunneling electron is accelerated by the laser field and away from the nuclei.
It can be probed at the detector
at the end of the laser pulse.
The corresponding physical process has been termed as above-threshold ionization (ATI) \cite{Agostini1979, Yang1993, Paulus1994, Lewenstein1995,Becker2002}.
The tunneling electron also has the chance to return to and recombine with the nuclei, with the emission of high-energy photon.
The corresponding process has been termed as high-order harmonic generation (HHG) \cite{McPherson1987,Huillier1991,Lewenstein1994}.
All of these steps of tunneling, acceleration and recombination can be manipulated by the laser field.

In comparison with linearly-polarized single-color or two-color laser fields \cite{Dudovich05,Zeng06,Mashiko07,YZ12},
orthogonally-polarized single-color \cite{Eckle1,Eckle2,Eckle3,Boge,Torlina,Quan,Undurti} or
two-color \cite{Kitzler2005,Brugnera2011,Kitzler2007,Shafir2009,Shafir2,Lein2013,chen2018}
laser fields provide richer manners for manipulating the electron motion.
For example, the elliptical laser field with high ellipticity is able to suppress the recombination.
It is also capable of resolving the direct and rescattering ATI electron trajectories \cite{Becker2002}.
These properties have been used to probe attosecond tunneling dynamics of the electron and
relevant probing procedures have been termed as attoclock \cite{Eckle1,Eckle2}.
On the other hand, the orthogonal two-color ($\omega,2\omega$) (OTC) laser field with a delay between these two colors
has shown the capability of controlling the long and short HHG electron trajectories \cite{Lewenstein1994}.
By adjusting this delay, HHG associated with only long or short electron trajectories can be selected \cite{Kitzler2005,Brugnera2011,chen2018}.
The OTC laser field can also map the momenta of photoelectrons,
which are born at different time regions in a laser cycle of the fundamental field,
into different regions of photoelectron momentum distribution (PMD) \cite{Xie}, somewhat similar to the elliptical case.
However, not as the case of elliptical laser field, where the offset angle  in PMD can be used as a convenient
tool to abstract the time information of photoelectron 
\cite{Eckle1,Eckle2},
the feature quantity in PMD of OTC which can be used in attosecond chronoscope is unclear so far.

In this paper, we study ionization of the He atom in strong OTC laser fields.
Through numerical solution of time-dependent Schr\"{o}dinger equation (TDSE) at diverse laser parameters, we identify
two characteristic structures in PMD.
The first one shows a rectangular structure located in the middle part of the whole PMD.
The second one presents a shoulder structure located at the top of the lower half plane of PMD.
By using a Coulomb-included strong-field model, which allows us to quantitatively describe the effect of the long-range Coulomb potential,
we reproduce the TDSE results, especially for these two typical structures. With this model, we are able to derive some simple formulae 
for the locations of these structures. The formulae show that these two structures arise from the Coulomb-induced ionization time lag,
which reflects the response time of the electronic wave function to a strong-laser induced tunneling-ionization event.
With the simple formulae, one can easily deduce the response time from the locations of these structures, especially from the shoulder-structure one.

\emph{TDSE}.-The Hamiltonian of He exposed to a strong laser field can be written as (in atomic units of $\hbar=e=m_{e}=1$)
${H}(t)={\mathbf{{p}}^2}/{2}+V(\mathbf{r})+\mathbf{E}(t)\cdot \mathbf{r}$.
Here,
$V(\textbf{r})$ is the Coulomb potential.
The OTC electric field $\mathbf{E}(t)$ used here has the  form of
$\mathbf{E}(t)=f(t)[\vec{\mathbf{e}}_{x}E_{x}(t)+\vec{\mathbf{e}}_{y}E_{y}(t)]$ with $E_{x}(t)=E_0\sin(\omega_0 t)$ and $E_{y}(t)=E_1\sin(2\omega_0 t+\phi_0)$.
Here, $E_1=\varepsilon E_0$ and $E_0$ ($E_1$) is the maximal laser amplitude relating to the peak intensity $I_x$ of $E_x(t)$ ($I_y$ of $E_y(t)$).
The term $\varepsilon$ is the ratio of  $E_1$ to $E_0$, $\omega_0$ is the laser frequency of $E_x(t)$ relating to the wavelength $\lambda_x$,
$\phi_0=\pi/2$ is the relative phase,  $f(t)$ is the envelope function,
and $\vec{\mathbf{e}}_{x}$($\vec{\mathbf{e}}_{y}$) is the unit vector along the $x(y)$ axis.
The TDSE
is solved numerically using spectral method \cite{Feit}.
Then we obtain PMD with both coherent and noncoherent treatments.
As to be shown below, the main features of PMD in OTC obtained with both treatments are similar,
but the coherent results show some complex interference structures and photon rings
which can not be well resolved in experiments for longer laser wavelengths \cite{Eckle1,Undurti}.
For simplicity, unless mentioned elsewhere, we focus our discussions on noncoherent three-dimensional (3D) results.
Numerical details can be found in \cite{Methods}.

\emph{TRCM}.-To analytically study  ionization of He in OTC fields,
we use the model developed in \cite{Chen2021}.
This model has been termed
as tunneling-response-classical-motion (TRCM) model. It arises from strong-field approximation (SFA)
\cite{Lewenstein1995} but considers the Coulomb effect \cite{MishaY,Goreslavski,yantm2010}.
This model first neglects the long-range Coulomb potential and
solves the SFA saddle-point equation
\begin{equation}
[\textbf{p}+\textbf{A}(t_s)]^2/2=-I_p
\end{equation}
to obtain the electron trajectory ($\textbf{p},t_0$).
Here, $\textbf{p}$ is the drift momentum of the photoelectron.
$\textbf{A}(t)$ is the vector potential of the electric field $\mathbf{E}(t)$.
The term $t_0$ denotes the tunneling-out time of the photoelectron.
It corresponds to the real part of the complex time $t_s=t_0+it_x$ that satisfies the saddle-point equation Eq. (1).
The trajectory ($\textbf{p},t_0$) agrees with the following mapping relation
\begin{equation}
\mathbf{p}\equiv\mathbf{p}(t_0)=\textbf{v}(t_{0})-\textbf{A}(t_{0}).
\end{equation}
Here, the term $\textbf{v}(t_{0})=\mathbf{p}+\textbf{A}(t_{0})$
denotes the exit velocity of the photoelectron at the exit position (i.e., the tunnel exit)  \cite{yantm2010}
$\mathbf{r}_0\equiv\mathbf{r}(t_0)=Re(\int^{t_0}_{t_0+it_{x}}[\mathbf{p}+\mathbf{A}(t')]dt')$.
The corresponding complex amplitude for the trajectory ($\textbf{p},t_0$)
can be expressed as $c(\textbf{p},t_0)\sim e^{b}$.
Here, $b$ is the imaginary part of the quasiclassical action
$S(\textbf{p},t_s)=\int_{t_s}\{{[\textbf{p}+\textbf{A}(t'})]^2/2+I_p\}dt'$ with $t_s=t_0+it_x$ \cite{Lewenstein1995}.

Then the TRCM further considers the Coulomb effect. It assumes that at the tunnel exit $\mathbf{r}(t_0)$, the tunneling electron
with the drift momentum $\textbf{p}$  is still located at a quasi-bound state which approximately agrees with  the virial theorem.
A small period of time $\tau$ is needed for the tunneling electron to evolve
from the quasi-bound state into an ionized state. Then it is free
at the time $t_i=t_0+\tau$ with the Coulomb-included drift momentum $\textbf{p}'$.
This time $\tau$ can be understood as the response time of the electron inside an atom to light in laser-induced photoelectric effects and is manifested as the Coulomb-induced ionization time lag in strong-field ionization \cite{Xie,Wang2020}.
The mapping between the drift momentum $\textbf{p}'$ and the ionization time $t_i$ in TRCM can be expressed as
\begin{equation}
\mathbf{p}'\equiv\mathbf{p}'(t_i)=\textbf{v}(t_{0})-\textbf{A}(t_{i}).
\end{equation}

\emph{Time Lag}.-According to the above assumptions in TRCM, at the tunneling-out time  $t_0$, the tunneling electron is still located
in a quasi-bound state $\psi_b$ with approximately agreeing with the virial theorem.
Specifically, the average potential energy of this state is
$\langle V(\mathbf{r})\rangle\approx V(\textbf{r}(t_0))$ and the average kinetic energy is
$\langle\textbf{v}^2/2\rangle=n_f\langle v_x^2/2\rangle\approx-V(\textbf{r}(t_0))/2$.
This state can be further approximately treated as a quasi particle
with the velocity $|\textbf{v}_{i}|=\sqrt{\langle v_x^2\rangle}\approx\sqrt{|V({\textbf{r}}(t_0))|/n_f}$
which has the direction opposite to the position vector $\textbf{r}(t_0)$.
A time lag $\tau$ is needed for the tunneling electron to acquire an impulse
$|\textbf{E}(t_0)|\tau=|\textbf{v}_{i}|$ from the laser field in order to overcome this Coulomb-induced velocity.
Then the lag $\tau$ can be evaluated with the expression of
\begin{equation}
\tau\approx\sqrt{|V({\textbf{r}}(t_0))|/n_f}/|\textbf{E}(t_0)|.
\end{equation}
Here, $n_{f}=2,3$ is the dimension of the system studied and the exit position
$\mathbf{r}(t_0)$ is determined by the saddle points of Eq. (1).
In single-active electron approximation, the potential  $V(\textbf{r})$ at the position $\mathbf{r}(t_0)$ can be considered to
has the form of $V(\textbf{r})\equiv V(r)=-Z/r$.
For comparisons with TDSE simulations, the effective charge $Z$ can be chosen as that used in simulations.
For comparisons with experiments, the value of $Z$ can be evaluated with $Z\approx\sqrt{2I_p}$.

\emph{PMDs}.-By assuming that for an arbitrary SFA electron trajectory ($\textbf{p},t_0$),
the Coulomb potential
 does not influence the corresponding complex amplitude  $c(\textbf{p},t_0)$,
we can obtain the TRCM amplitude  $c(\textbf{p}',t_i)$ for Coulomb-included electron trajectory ($\textbf{p}',t_i$) directly from the SFA one
with $c(\textbf{p}',t_i)\equiv c(\textbf{p},t_0)$ at $\tau\approx\sqrt{|V({\textbf{r}}(t_0))|/n_f}/|\textbf{E}(t_0)|$.
This TRCM therefore allows the analytical evaluation of the Coulomb-included PMD.

\emph{MPR}.-Next, we discuss the most probable route (MPR)  \cite{chen2021} for electron trajectories in the present OTC field.
The MPR corresponds to the momentum with the maximal amplitude in PMD, which is easier to identify.

Clearly, for $\omega_0 t_0=\pi/2$ or $3\pi/2$, both the fundamental field $E_x(t_0)=E_0\sin(\omega_0 t_0)$
and the second-harmonic field $E_y(t_0)=E_1\sin(2\omega_0 t_0+\pi/2)$ have the maximal amplitudes of $|E_0|$ and $|E_1|$, respectively.
Without loss of generality, we consider the case of $\omega_0 t_0=\pi/2$.
For this case,  $A_x(t_0)=A_y(t_0)=0$. In addition, our simulations show $v_x(t_0)\approx v_y(t_0)\approx0$.
Therefore, by Eq. (2), we have $\textbf{p}\equiv\textbf{p}(t_0)\approx0$. That is, for the MPR with $\omega_0 t_0=\pi/2$,
the SFA predicts zero momentum that has the maximal amplitude. By Eq. (3) of TRCM, we also have
\begin{equation}
p'_x\approx-{A}_x(t_{i})\approx E_0\tau;{}{} p'_y\approx-{A}_y(t_{i})\approx -E_1\tau.
\end{equation}
Similarly, for the case of  $\omega_0 t_0=3\pi/2$, we have $p'_x\approx -E_0\tau, p'_y\approx -E_1\tau$.
Here, the lag related to the MPR is $\tau\approx\sqrt{|V({\textbf{r}}(t_0))|/n_f}/\sqrt{E_0^2+E_1^2}$.
These expressions imply that when the Coulomb effect is considered, the MPR related to zero momentum in SFA is shifted
to nonzero momenta $p'_x\approx\pm E_0\tau,p'_y\approx -E_1\tau$ in TRCM.

\begin{figure}[t]
\begin{center}
\rotatebox{0}{\resizebox *{8.5cm}{6cm} {\includegraphics {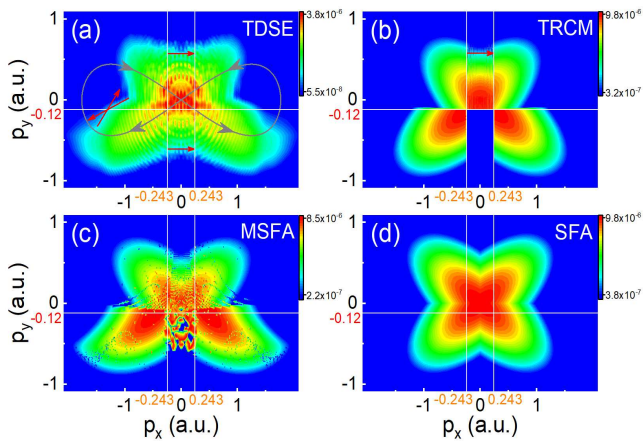}}}
\end{center}
\caption{Comparisons of PMD obtained with TDSE (a), TRCM (b), MSFA (c), and SFA (d) in OTC.
The prediction of SM (gray line) is also plotted in (a).
Laser  parameters used are $I_x=5\times10^{14}$ W/cm$^{2}$, $\lambda_{x}=800$ nm and $\varepsilon=0.5$. The $\log_{10}$ scale is used here.
In each panel,
the vertical lines indicate the axes of $p'_x=\pm E_0\tau$ and the horizontal one indicates the axis of $p'_y=-E_1\tau$ of Eq. (5).
Relevant values of $p'_x$ and $p'_y$ are denoted with the color numbers.
The intersection of the oblique arrows in (a) indicates the position of the shoulder structure
and the horizontal arrows indicate the width of the rectangular structure.
}
\label{fig:g1}
\end{figure}

\emph{Results and discussions}.-In Fig. 1, we show the PMDs of He in OTC calculated with different methods.
The TDSE results in Fig. 1(a) show a butterfly shape with
two smaller upper wings and two larger lower wings.
In particular, the PMD of TDSE presents two typical structures.
The first one is a rectangular-like structure  and is located in the middle part of the whole PMD.
It is like the body of the butterfly and the width of this structure can be identified
through the relatively flat part between the two upper or lower wings as indicated by the horizontal arrows.
The left and the right boundaries of this structure approximately agree with the relation of $p'_x=\pm E_0\tau$ of Eq. (5),
as indicated by these two vertical lines.

The second one is a shoulder-like structure and is located at the top of the lower half plane of the PMD.
It has a small slope as indicated by the downward oblique arrow.
Below the shoulder structure,
the fringe of the distribution shows the large gradient, as indicated by the upward oblique arrow.
The intersection of these two arrows indicates the shoulder position
which agrees with the relation of $p'_y=-E_1\tau$ of Eq. (5), as indicated by the horizontal line.
The main features of PMD of TDSE are well reproduced by the TRCM, as shown in Fig. 1(b).
Particularly, the positions of these two typical structures predicted by TRCM are in quantitative agreement with the TDSE ones.
For clarity, we also indicate these positions with the vertical and horizontal lines in Fig. 1(b).

By comparison, the PMD obtained with the Coulomb-modified SFA (MSFA) model \cite{Xie}, is similar to the TRCM one, but
the positions of these two typical structures differ remarkably from the TRCM predictions of $p'_x=\pm E_0\tau$ and $p'_y=-E_1\tau$,
as indicated by the lines. The rectangular structure predicted by the MSFA is narrower than the TRCM one and the position
of the shoulder structure is near to the axis of $p_y=0$.
Different from the TRCM which mainly considers the Coulomb effect near the nucleus with the virial theorem,
the MSFA considers the Coulomb effect through numerical solution of the Newton equation including both
the electric-field force and the Coulomb force for each SFA electron trajectories after the tunneling electron exits the barrier.

\begin{figure}[t]
\begin{center}
\rotatebox{0}{\resizebox *{8.5cm}{8cm} {\includegraphics {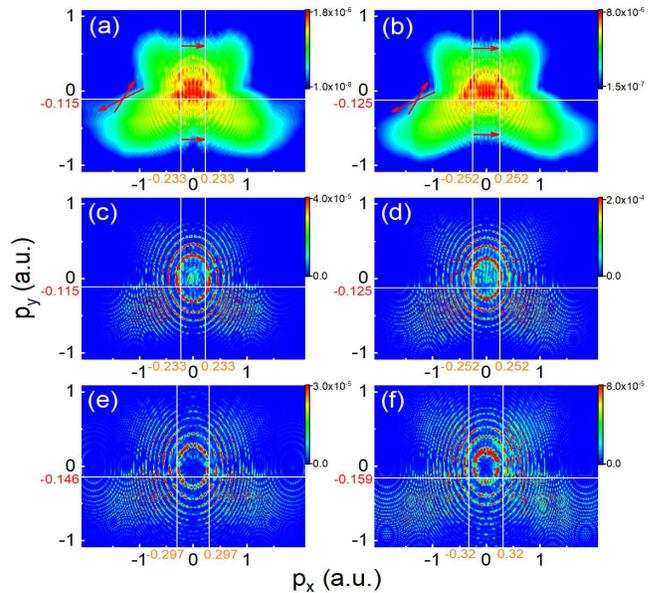}}}
\end{center}
\caption{PMDs of He in OTC obtained with TDSE  for  $I_x=4\times10^{14}$W/cm$^{2}$ (first column) and $I_x=6\times10^{14}$W/cm$^{2}$ (second) in 3D noncoherent (first row) and coherent cases (second) and 2D coherent cases (third).
Other laser parameters used are $\lambda_x=800$ nm and $\varepsilon=0.5$.
The $\log_{10}$ scale is used in the first row.
The lines and arrows are as in Fig. 1.
}
\label{fig:g2}
\end{figure}

Let us further understand the origin of the butterfly structure of PMD in OTC with the TRCM.
The PMD obtained with SFA where the Coulomb effect is not considered is presented in Fig. 1(d).
The PMD of SFA shows a high symmetry with respect to both the axes of $p_x=0$ and $p_y=0$.
In addition, the distributions of SFA along these two axes both have large amplitudes.
When the Coulomb effect is considered with TRCM, the mapping of Eq. (3) begins to work.
In comparison with Eq. (2) of SFA predictions, Eq. (3) shows that
the Coulomb effect induces a time lag $\tau$ of the ionization time $t_i$ relative to
the tunneling-out time $t_0$. For $p_y\geq0$, this lag results in the shift of the whole right (left)
part of SFA distribution with $p_x\geq0$ ($p_x\leq0$)
towards the left (right), and the situation reverses for $p_y<0$. In addition,
this lag also induces the shift of the entire upper part of SFA distribution
towards the lower region.
More specifically, this shift follows the direction of the gray arrow for the SM predictions of $\textbf{p}=-\textbf{A}(t)$ plotted in Fig. 1(a).
The left and right offsets of the axis $p_x=0$ forms the boundaries of the rectangular structure
around $p'_x\approx\pm E_0\tau$ and the downward offset of the axis $p_y=0$ forms the shoulder structure.
Our analyses  show that the TRCM predictions of momenta $\textbf{p}'$ are similar for SFA momenta $\textbf{p}$
along the axis of $p_x=0$ or $p_y=0$.
Therefore, if the  structures can be quantitatively identified in PMD, one may obtain the lag $\tau$ from
the positions of the structures with Eq. (5).
Then the ionization time (the birth time) $t_i=t_0+\tau$ of the tunneling electron can be determined.
However, the boundary of the rectangular structure is not regular enough due to the interference
between the shifted distributions from different parts.
By comparison, the shoulder structure is more regular and the position of this structure is easier to determine.
To the best of our knowledge, this is the first quantitative explanation on the origin of the butterfly structure of PMD in OTC.

\begin{figure}[t]
\begin{center}
\rotatebox{0}{\resizebox *{8.5cm}{8cm} {\includegraphics {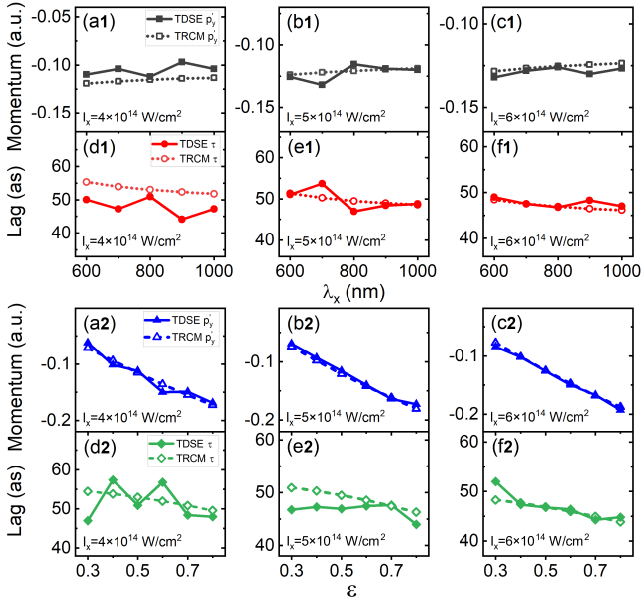}}}
\end{center}
\caption{Comparisons of the shoulder position $p'_y$  in PMD (first and third rows) and the corresponding time lag
$\tau$ (second and fourth rows) predicted by TDSE and TRCM
for different $\lambda_x$ at $\varepsilon=0.5$ (a1-f1)  and
different $\varepsilon$ at $\lambda_x=800$ nm (a2-f2). The laser intensities $I_x$ used are as shown.
The TDSE shoulder position is obtained 
as in Fig. 1
and the TRCM one is obtained with $p'_y=-E_1\tau$ of Eq. (5). The corresponding lags are obtained
through  $\tau\approx\sqrt{|V({\textbf{r}}(t_0))|/n_f}/\sqrt{E_0^2+E_1^2}$ of TRCM predictions and
through $\tau=|p'_y|/E_1$ with $|p'_y|$ being TDSE predictions of the shoulder position.
}
\label{fig:g3}
\end{figure}

In Fig. 2, we show PMDs of He in OTC calculated by TDSE with both noncoherent and coherent treatments at different laser intensities.
For comparison, coherent results with two-dimensional (2D) simulations are also shown.
One can observe that for all of the noncoherent and coherent results in 3D and 2D cases, these two typical structures can be well resolved.
The positions of these structure are also in agreement with the TRCM predictions
as indicated by the solid lines of $p'_x=\pm E_0\tau$ and $p'_y=-E_1\tau$ of Eq. (5).
It is worth noting that although the coherent results show some complex interference structures,
the shoulder positions can also be clearly identified from the flat contour of this structure around $p_x=\pm1$ a.u..
Here, no trajectories  contributes to the distribution higher than the contour in TRCM (also see Fig. 1(b)).
In addition, the position in 2D cases associated with $n_f=2$ in Eq. (4)
is larger than in 3D cases with $n_f=3$.
Thirdly, for the present cases, the shoulder position is not very sensitive to the laser intensity.

As the shoulder position can be more easily determined  (as shown in Fig. 1 and Fig. 2),
in Fig. 3, we compare the TDSE predictions of this position to the theoretical ones
at various laser parameters.

In the first row of Fig. 3, we show relevant comparisons at different laser intensities and wavelengths.
Firstly, the TDSE and theoretical results agree well with each other for different cases, with an average error smaller
than $10\%$. Secondly,  for a fixed intensity, the position changes slowly with increasing the wavelength.
Thirdly, for a fixed wavelength, the position also shows a slow increase when the laser intensity increases.
In the second row of Fig. 3, we show the corresponding time lag $\tau$ obtained with TRCM and TDSE.
One can observe that for the present cases, the theoretical prediction of $\tau$ is located at a range of about $45$ to $55$ attoseconds and
decreases slowly both for increasing the laser wavelength and intensity.
The TDSE results agree well with the theoretical ones with an average difference smaller than $5$ attoseconds.
This agreement is more remarkable for higher laser intensities and longer  wavelengths $\lambda_x$ on the whole.

In the third and the fourth rows of Fig. 3, we show further comparisons
for different ratios $\varepsilon$. It can be observed from the third row,
the TDSE and TRCM predictions of the shoulder position are in good agreement with each other.
In addition, for a fixed laser intensity, the position increases remarkably when increasing the ratio $\varepsilon$,
with the position  $p'_y$ changing from $p'_y\approx-0.07$ a.u. at $\varepsilon=0.3$ to $p'_y\approx-0.18$ a.u. at $\varepsilon=0.8$.
By comparison, the corresponding time lag $\tau$ in the fourth row of Fig. 3 is located at a smaller range as increasing $\varepsilon$.
This range is about $50$ to $55$ attoseconds in Fig. 3(d2), $46$ to $51$ attoseconds in Fig. 3(e2) and $43$ to $48$ attoseconds in Fig. 3(f2).
The TDSE and theoretical predictions of $\tau$ are still well consistent with each other here, with a difference smaller than $5$ attoseconds.
This difference is smaller for cases of higher laser intensities and larger $\varepsilon$,
suggesting the preferred laser parameters in experiments.

\emph{Conclusion}.-In summary, we have studied ionization of He in strong OTC laser fields.
The PMDs of He in OTC show two typical structures, i.e., the rectangular one and the shoulder one related to the most probable route.
The locations of these two structures encode
the response time of electron inside an atom to light in laser induced tunneling ionization.
This time can be easily extracted from the position of the shoulder structure that can be well identified in PMD.
The OTC laser field has wide applications in manipulating the electron motion in both ATI and HHG.
The two-color attosecond chronoscope proposed here provides a simple tool for timing the birth of electron in relevant OTC-based  manipulations.

This work was supported by the National Natural Science Foundation of China (Grant Nos. 12174239, 11904072).

\appendix

\section{Numerical Methods}

In the length gauge and single-active electron approximation,
the Hamiltonian of the He atom interacting with a strong laser field can be written as (in atomic units of $\hbar=e=m_{e}=1$)
\begin{equation}
{H}(t)=H_{0}+\mathbf{E}(t)\cdot \mathbf{r}
\end{equation}
Here, $H_{0}={\mathbf{{p}}^2}/{2}+V(\mathbf{r})$ is the field-free Hamiltonian and
$V(\textbf{r})=-Z/\sqrt{r^2+\xi}$
is the Coulomb potential with the effective charge $Z$ and the soft-core parameter $\xi$.
In three-dimensional (3D) cases, we have used the
parameters of $Z=1.34$ and $\xi=0.071$.
With these parameters, the ionization potential of He reproduced is $I_p=0.9$ a.u..
The term $\mathbf{E}(t)$ denotes the electric field of the OTC laser field as introduced in the main text.

We use trapezoidally shaped laser pulses with a total duration of fifteen cycles,
which are linearly turned on and off for three optical cycles, and then kept at a constant intensity for nine
additional cycles. The TDSE of $i\dot{\Psi}(\textbf{r},t)=$H$(t)\Psi(\textbf{r},t)$ is solved numerically
using the spectral method \cite{Feit} with a time step of $\triangle t=0.05$ a.u..
We have used a grid size of  $L_x\times L_y\times L_z=358.4\times 358.4\times 51.2$ a.u.
with $\triangle x=\triangle y=0.7$ a.u. and $\triangle z=0.8$ a.u..
The numerical convergence is checked by using a finer grid.

In order to avoid the reflection of the electron wave packet from the boundary and obtain the momentum space wavefunction, the coordinate
space is split into the inner and the outer regions with ${\Psi}(\textbf{r},t)={\Psi}_{in}(\textbf{r},t)+{\Psi}_{out}(\textbf{r},t)$,
by multiplying a mask function of $F(\mathbf{r})$. The mask function has the form of $F(\mathbf{r})=F_1(x)F_2(y)F_3(z)$. Here,
$F_1(x)=\cos^{1/8}[\pi(|{x}|-r_x)/(L_x-2r_x)]$ for $|{x}|\geq r_x$ and $F_1(x)=1$  for $|{x}|< r_x$. $r_x=134$ a.u. is the absorbing boundary.
The form of $F_2(y)$ is similar to $F_1(x)$.
The expression of $F_3(z)$ is also similar to  $F_1(x)$ but the absorbing boundary along the $z$ direction is $r_z=19.2$ a.u..
In the inner region, the wave function ${\Psi}_{in}(\textbf{r},t)$ is propagated with the complete Hamiltonian $H(t)$. In the outer region, the time evolution of the wave function ${\Psi}_{out}(\textbf{r},t)$ is carried out in momentum space with the Hamiltonian of the free electron in the laser field.
The mask function is applied at each time  interval  of 0.5 a.u. and the obtained new fractions of the outer wave function are
noncoherently (coherently) added
to the momentum-space wave function $\tilde{{\Psi}}_{out}(\textbf{r},t)$ from which we obtain the noncoherent (coherent) PMD.

For comparison with 3D results, we have also performed two-dimensional (2D) simulations.
In 2D cases, we have used the
parameters of $Z=1.45$  and $\xi=0.5$ for the potential $V(\textbf{r})=-Z/\sqrt{r^2+\xi}$.
The grid size used in 2D cases is $L_x\times L_y=409.6\times 409.6$ a.u. with space steps
of $\triangle x=\triangle y=0.4$ a.u.. The mask function used is $F(\mathbf{r})=F_1(x)F_2(y)$.
The forms of $F_1(x)$ and  $F_2(y)$ are similar to 3D cases with the absorbing boundary of $r_x=r_y=150$ a.u..

\end{document}